\documentclass[]{spie}  %>>> use for US letter paper
\usepackage{amsmath,amsfonts,amssymb}
\usepackage{graphicx}
\usepackage[colorlinks=true, allcolors=blue]{hyperref}

\title{The path to detecting extraterrestrial life with astrophotonics}

\author[a]{Nemanja Jovanovic} %ORCID 0000-0001-5213-6207
\author[a]{Yinzi Xin} %ORCID 0000-0002-6171-9081
\author[b]{Michael P. Fitzgerald} %ORCID 0000-0002-0176-8973
\author[c,d]{Olivier Guyon}
\author[e,f]{Peter Tuthill}
\author[e,f]{Barnaby Norris} %ORCID 0000-0002-8352-7515
\author[a]{Pradip Gatkine} %ORCID 0000-0002-1955-2230
\author[a]{Greg Sercel}
\author[a]{Svarun Soda}
\author[b]{Yoo Jung Kim} %0000-0003-1392-0845
\author[b]{Jonathan Lin} %ORCID 0000-0001-8542-3317
\author[e,f]{Sergio Leon-Saval} %ORCID 0000-0002-5606-3874
\author[h]{Rodrigo Amezcua-Correa}
\author[h]{Stephanos Yerolatsitis}
\author[c]{Julien Lozi}
\author[c,d]{Sébastien Vievard} %ORCID 0000-0003-4018-2569
\author[e,f]{Chris Betters}
\author[i]{Steph Sallum}
\author[i]{Daniel Levinstein}
\author[a,j]{Dimitri Mawet} %ORCID 0000-0002-8895-4735
\author[j]{Jeffrey Jewell} 
\author[j]{J. Kent Wallace} 
\author[k]{Nick Cvetojevic} 

\affil[a]{Department of Astronomy, California Institute of Technology, 1200 E California Blvd, Pasadena, CA, 91125, USA}
\affil[b]{Department of Physics \& Astronomy, 430 Portola Plaza, University of California, Los Angeles, CA 90095, USA}
\affil[c]{National Astronomical Observatory of Japan, Subaru Telescope, 650 North Aohoku Place, Hilo, HI 96720, U.S.A.}
\affil[d]{Astrobiology Center, 2-21-1, Osawa, Mitaka, Tokyo, 181-8588, Japan}
\affil[e]{Sydney Astrophotonic Instrumentation Laboratory, School of Physics, The University of Sydney, Sydney, NSW 2006, Australia}
\affil[f]{Sydney Institute for Astronomy, School of Physics, Physics Road, The University of Sydney, NSW 2006, Australia}
\affil[h]{The College of Optics and Photonics, University of Central Florida, 4304 Scorpius St, Orlando, FL 32816}
\affil[i]{Department of Physics \& Astronomy, University of California, Irvine, 4129 Frederick Reines Hall, Irvine, CA 92697 USA}
\affil[j]{Jet Propulsion Laboratory, California Institute of Technology, 4800 Oak Grove Drive, Pasadena, CA, 91109, USA}
\affil[k]{Laboratoire Lagrange, Observatoire de la C\^{o}te d'Azur, Universit\'{e} C\^{o}te d'Azur, 06304 Nice, France}

\authorinfo{Further author information: (Send correspondence to N.J)\\N.J: E-mail: nem@caltech.edu}

\begin{document} 
\maketitle

\begin{abstract}
Astrophysical research into exoplanets has delivered thousands of confirmed planets orbiting distant stars. These planets span a wide ranges of size and composition, with diversity also being the hallmark of system configurations, the great majority of which do not resemble our own solar system. Unfortunately, only a handful of the known planets have been characterized spectroscopically thus far, leaving a gaping void in our understanding of planetary formation processes and planetary types. To make progress, astronomers studying exoplanets will need new and innovative technical solutions. Astrophotonics -- an emerging field focused on the application of photonic technologies to observational astronomy -- provides one promising avenue forward. In this paper we discuss various astrophotonic technologies that could aid in the detection and subsequent characterization of planets and in particular themes leading towards the detection of extraterrestrial life. 
\end{abstract}


\section{INTRODUCTION}
\label{sec:intro}  % \label{} allows reference to this section
\begin{figure} [ht]
\begin{center}
\includegraphics[width=0.7\textwidth]{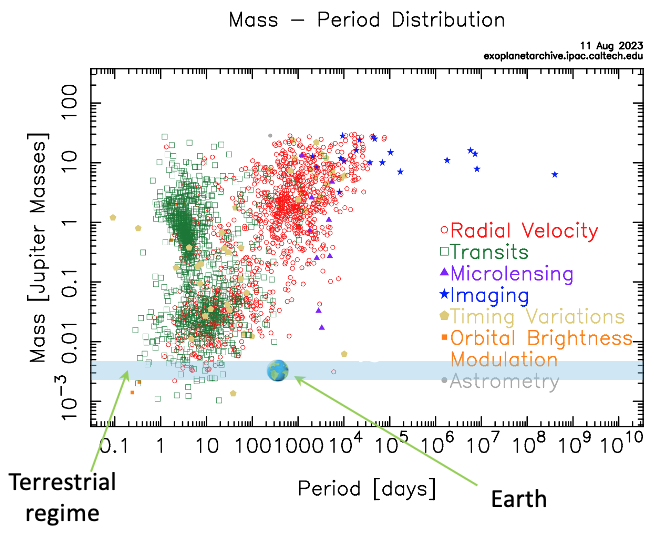}
\end{center}
\caption[example] 
   { \label{fig:detected_planets} 
5400+ confirmed exoplanets as of the 11$^{th}$ of August 2023 on a Jupiter mass vs orbital period plot. The Earth is shown on the plot to indicate where an Earth-like planet around a sun-like star would appear. The blue region highlights the terrestrial regime. Very few Earth mass/size planets have been detected thus far. Only a handful are in the habitable zone of their host star. Data taken from exoplanetsarchive.ipac.caltech.edu }
\end{figure}

There have been over 5400 exoplanets confirmed to date. Figure~\ref{fig:detected_planets} shows the mass vs orbital period distribution of the known planets. Data is color coded to highlight the detection technique used. For the vast bulk of this population of planets, very little is known. Spectroscopy of the exoplanet atmosphere is critical to revealing a wealth of information (composition and abundance, spin rate, weather patterns, etc). The field of exoplanetary sciences is now focusing efforts on characterization of these systems by for example providing direct exoplanet spectoscopy~\cite{Snellen2015,Wang2017}. By understanding more about the known exoplanets, we can refine planetary formation and evolution models and better understand where life is likely to exist.

The blue region in Fig.~\ref{fig:detected_planets} highlights the terrestrial planet regime (planets with Earth sizes/masses). Only a handful of terrestrial mass planets have thus far been detected. The Earth is overlaid highlighting that we don't currently know of an Earth/Sun analog, which would constitute a primary candidate in the search for life. Of the planets that have been detected in this regime, very few are in the habitable zone of their host planets. To detect extraterrestrial life, we must first detect more terrestrial like planets in the habitable zones of their host stars. In particular, finding planets around solar type stars (G stars) is critical for studying systems similar to our own to understand our place in the Universe.  

\section{Detecting terrestrial planets}
The transit and radial velocity techniques have detected the bulk of the known exoplanets thus far. Although both techniques will continue to add more terrestrial candidates around M type stars, the radial velocity technique is the most likely technique to detect Earth-like planets around G type stars, needed to form the target list for the Habitable Worlds Observatory (HWO) mission. This will require measuring a velocity change in the speed of light of $<$50 cm/s and more likely 10-20 cm/s (1 part in $3\times10^{9}$)~\cite{Fischer_2016}. And if the Earth-like planet is in the habitable zone, this velocity change would occur over half a year, so the rate of change is extremely small. For this reason this endeavour is known as extreme precision radial velocity (EPRV).    

\subsection{Laser Frequency Combs}
To be able to measure such a small velocity change, it is critical to be able to calibrate out any instrument drifts to better than 1 cm/s. This requires an extremely stable spectral calibration source which is known as the laser frequency comb (LFC). An LFC emits a series of ultra-stable, uniformly spaced lines across a broad spectrum that can be used to calibrate an instrument over many years and even decades.

There are numerous ways to generate the comb. A mode-locked laser was originally used for this application. It generates an extremely densely packed comb (comb line spacing's are in the MHz)~\cite{sinclair_ACO_2015}. At such small line spacing's an R~100,000 spectrograph, typically applied to EPRV can not separate the lines out. Fabry-Perot Etalon filter cavities are therefore used downstream to remove the bulk of the lines and establish line spacing's of 10-30 GHz, which are better suited for astronomical spectrometers~\cite{Li_ALF_2008,Ycas_DOS_2012}.  

Electro-optic combs rely on electo-optic modulators to generate side bands on a continuous-wave laser line~\cite{Obrzud_BNI_2018}. This allows the comb to be formed with the adequate line spacing from the outset. The line spacing can be stabilized by locking it to a clock reference for example. 

Finally, LFCs have also been realized on a chip~\cite{obrzud_AMC_2019}. These systems are still not quite turn key, but once they are matured may be applicable to many applications that have tight space constraints, like those in space. 

The preliminary spectrum of the LFC is then sent through a non-linear medium, such as a highly nonlinear fiber or crystal to broaden it. In this way its possible to realize combs that span more than 1 octave. 

LFCs can be locked to clock references and also to one another allowing for the frequency uncertainty to be easily reduced to well below those needed for EPRV. These locking loops also ensure that the frequencies of the lines can be maintained for decades~\cite{Hansch_NLP_2006}, which will be critical to detecting Earth like planets around G stars. 

\subsection{Spectral flatteners-on-a-chip}
The amplitude across an LFC can vary by orders of magnitude. The precision of the wavelength solution scales with the square root of the number of lines used to derive it, and the signal-to-noise of a line scales with the square root of the number of photons in a line. To optimize the wavelength solution the lines would have uniform amplitude with high signal-to-noise ratio in a given exposure time. 

To achieve this with an LFC, the spectrum must first be flattened. Traditionally a flattener consists of a bulk optic set up that collimates the light out of a single mode fiber (SMF), disperses it with a diffraction grating and bounces it off a spatial light modulator (SLM) which can control the reflected amplitude of the beam before it is spectrally recombined and injected into another SMF to be sent to the science instrument~\cite{Probst_SPS_2013}. A compact spectrometer is also needed to measure the actual spectrum to drive the SLM to flatten the output. This system is bulky, costly and unstable. 

Recently on-chip flatteners have been investigated~\cite{Jovanovic_FLF_2022}. These consist of the same basic elements on a chip. An arrayed-waveguide grating (AWG) is used to disperse the light into a series of discrete output channels before light is sent through Mach-Zehnder interferometers (MZI) which can be used to adjust the amplitude of the channel by thermally tuning the phase of one of the arms of the MZI before a thermo-optic phase modulator is used to re-phase all the spectral channels before a final recombination of the entire spectrum in another AWG. The prototype device developed operated on a single polarization, over a 400 nm range in the astronomical H band and offered $\sim38$ dB of amplitude tuning. 

Spectral flatteners on a chip are useful for ground based LFCs but will be critical to any future mission that requires spectral shaping in flight.

LFCs combined with spectral flatteners could play a key role in enabling state-of-the-art spectrometer calibration and enable terrestrial planet detection over the next decade.

\section{Characterizing Terrestrial Planets}
\label{sec:title}
The most effective way to characterize an exoplanet is to collect a spectrum. A spectrum can indicate the presence of molecules needed to sustain life, like water and oxygen, as well as ozone which might protect the planet from UV and greenhouse gases like carbon monoxide, carbon dioxide and/or methane. This is the most effective way to determine if a planet could harbor life. 

To take a spectrum of an Earth-like planet in the habitable zone we must first reduce the glare from the host star, which could be $10^{10}\times$ brighter. Given the small angular separation to the orbital distance of the habitable zone planet, this will require a large telescope ($>6$ m sized aperture) and advanced high contrast imaging techniques. These consist of wavefront sensing and control to eliminate aberrations and improve contrast, followed by starlight suppression from for example a coronagraph. Although there has been tremendous progress in coronagraph performance over the past few decades, these systems have still not achieved the requirement of $10^{10}$ contrast across a 10-20$\%$ bandwidth needed to take a spectrum.

Photonics can be used to support the wavefront sensing and control and starlight suppression aspects of the high contrast imaging system as well as for the science instrument as outlined below. 

\subsection{Photonic Lantern Wavefront Sensing \& Control}
Photonic technologies offer the ability to coherently mix light, necessary for generating signals for wavefront sensing. Although there are several possible approaches, photonic lanterns provide a convenient avenue. Photonic lanterns are waveguide devices that convert a multimoded input into several single mode outputs via an adiabatic transition. As long as the number modes at the single mode end is greater than the number of modes in the multimode end, the transition will be efficient~\cite{LeonSaval_PL_2013}. 

When a lantern is placed in a focal plane, the input beam is coupled amongst the modes the lantern supports and traverses the transition to the output array of SMFs. Owing to their few moded nature (3, 6, 19, 61 mode counts are typical), lanterns have a greater collecting efficiency as compared to SMFs. At the output, the flux distribution across the ports uniquely encodes the information about the complex field of the incident beam. Therefore, a lantern could be placed in a focal plane to collect light for a downstream instrument and provide the ability to do wavefront sensing. Focal plane wavefront sensors of this nature can eliminate non-common path and chromatic errors. 

It is possible to use a neural net to map the input electric field to output intensity distributions to maximize the dynamic range of the lantern~\cite{Norris_AAP_2020} and use this mapping to reconstruct the input complex field. Operating the lantern in the linear regime has also been modelled~\cite{Lin_FPW_2022} and recently demonstrated on the SCExAO testbed~\cite{Jovanovic_SCE_2015}. This test was done around 1550 nm with a 19 port photonic lantern and demonstrated the successful closed loop control of 5+ of the low order Zernike modes off-sky. The outcome of this work is discussed in another paper in the same conference~\cite{lin_DLN_2023}. 

An interesting prospect is to use a hybrid lantern - a lantern that can transport light injected into the LP01 mode to one isolated output, while the other ports consist of a combination of light from each of the modes (i.e. are coherently mixed)~\cite{Norris_OBS_22}. This concept allows for light to be routed directly to a science instrument while also providing on-board wavefront sensing which can allow for improved coupling to the lantern. It was recently proposed and simulated but has not been demonstrated.  

\subsection{Photonic Nulling}
Starlight cancellation is extremely important to reduce the photon noise, which would otherwise dominate a spectrum collected from the planet. There are numerous technologies/approaches that could be used to suppress starlight using photonics. 

GLINT is a photonic instrument in operation on the SCExAO testbed at Subaru telescope~\cite{Norris_FOS_2019}. It relies on segmenting the pupil and injecting each of the beamlets into a pupil remapping chip realized with ultrafast laser inscription. The beams are routed in 3D inside the photonic component and tapped via splitters for the purpose of photometric monitoring. Flux in the main channel is combined pairwise in photonic couplers with beams from other parts of the pupil. Carefully arranging the relative phases between combined beams generates nulled signals. For our case this can be done across many baselines simultaneously. To improve Fourier coverage, nulls on numerous baselines need to be obtained simultaneously. Scaling the number of input channels is possible with this approach~\cite{Antoine_SPB_2021}, but getting achromatic nulls across multiple baselines at once remains challenging. Solutions are being proposed to design more achromatic circuits\cite{Martinod_APT_2021}. 

Photonic lanterns can also be used for nulling. Specifically, mode-selective lanterns - lanterns that map LP modes to unique single-mode outputs naturally provide a nulling capability. For a 6 port lantern, LP11a, LP11b, LP21a, and LP21b all have phase inversions on axis prohibiting light in a pure even-symmetry mode to couple into them~\cite{Xin_EDC_2022,Tuthill_NIH_2022}. Therefore, at the output SMFs corresponding to these modes, the starlight is suppressed to some extent while planet light is coupled. This has recently been demonstrated in the laboratory with a 6 port mode selective lantern operating around 1550 nm~\cite{Xin_LCM_2023}. Preliminary results show monochromatic and 10\% bandwidth polychromatic null depths ranging from 10$^{-3}$-10$^{-2}$ across the 4 nulled outputs, which were limited by the finite cross-talk between the modes in the device. Next steps include developing lower cross-talk devices. Nonetheless, this passive component provides a simple avenue to improve contrast between 0.5-2~$\lambda/D$. A possible extension of this concept would be to realize a hybrid lantern that allows the two LP11 and two LP21 modes to be separated into unique SMF outputs, while keeping all other modes coherently mixed together. If this were possible, it would allow for only the channels used in nulling to be separated and used for this application, while all of the rest of the light from the planet would be used for focal plane wavefront control. This could improve the contrast and stabilize it as well. 

Similar to the GLINT concept, its also possible to either segment the focal plane into an integral field unit that subtends the inner 2$\lambda/D$ or use a standard lantern to collect the light. At the output, the beams from the various parts of the focal plane, or the ports of the lantern can be interfered pair wise using directional or multi-mode interference couplers. Indeed, several stages of beam combination with fine path length adjustments could be used to reduce any light leak from an upstream coupler due to for example imperfections in the coupler or difficulties in phasing, to get to deeper nulls~\cite{Wilkes_6dB_2016}. This circuit could also be combined with the mode-selective nuller above as a second stage of nulling. For a mission like the HWO, the contrast is extreme and photon rates in the final stage will be low, so it may be difficult to generate sufficient signals to phase the circuit appropriately. This is something that would need to be investigated. Kernel nulling self-calibration can also be applied to analyze the output of such circuits and will also provide a boost in the contrast~\cite{cvetojevic_TBS_20223}.

\subsection{Photonic Spectroscopy}
As outlined above, collecting a spectrum of the planet itself is critical to confirming that the planet can host life. Photonic spectrometers in the form of AWGs form an ideal solution as they are 1) compact and can be flown on the HWO, 2) consist of a single-monolithic component with no moving parts that can be readily thermally stabilized, and 3) offer a discretized output that can be routed to a detector located elsewhere.

At such extreme contrasts, even on 6-m apertures, the photon rates from the exoplanets are very low. Therefore, lower resolving powers ranging from R$\sim 50$ and up to $1000$ are being considered for the NIR characterization channel\cite{gatkine2017arrayed}. Typical commercial AWGs operate at R$\sim$7000 \cite{cvetojevic2012first}, so this represents a reduction in resolution. However, larger bandwidths than those typically used in telecommunications will be needed. A minimum bandwidth of $10\%$ (150~nm at 1550 nm) will be needed and more likely even broader bandwidths. When operating at low resolutions its possible to indeed broaden the free spectral range (FSR) as well as the bandwidth of the device. 

We have recently developed several low resolution devices between Caltech/JPL in both SiN and Silica photonics. The SiN device was optimized for a FSR of 500 nm while the Silica device was optimized for a FSR of 220 nm. Both devices were designed for a channel spacing of 8 nm at 1600 nm, corresponding to an R$\sim$200. The devices have recently been characterized and the results will be presented in detail in an upcoming publication. But its worth noting that the Silica device had an end-to-end efficiency $>70\%$ including fiber coupling to and from the chip. This demonstrates the viability of optimizing AWGs for exo-Earth characterization on the future HWO mission.

\section{Characterization Instrument Architecture}
\begin{figure} [ht]
\begin{center}
\includegraphics[width=0.99\textwidth]{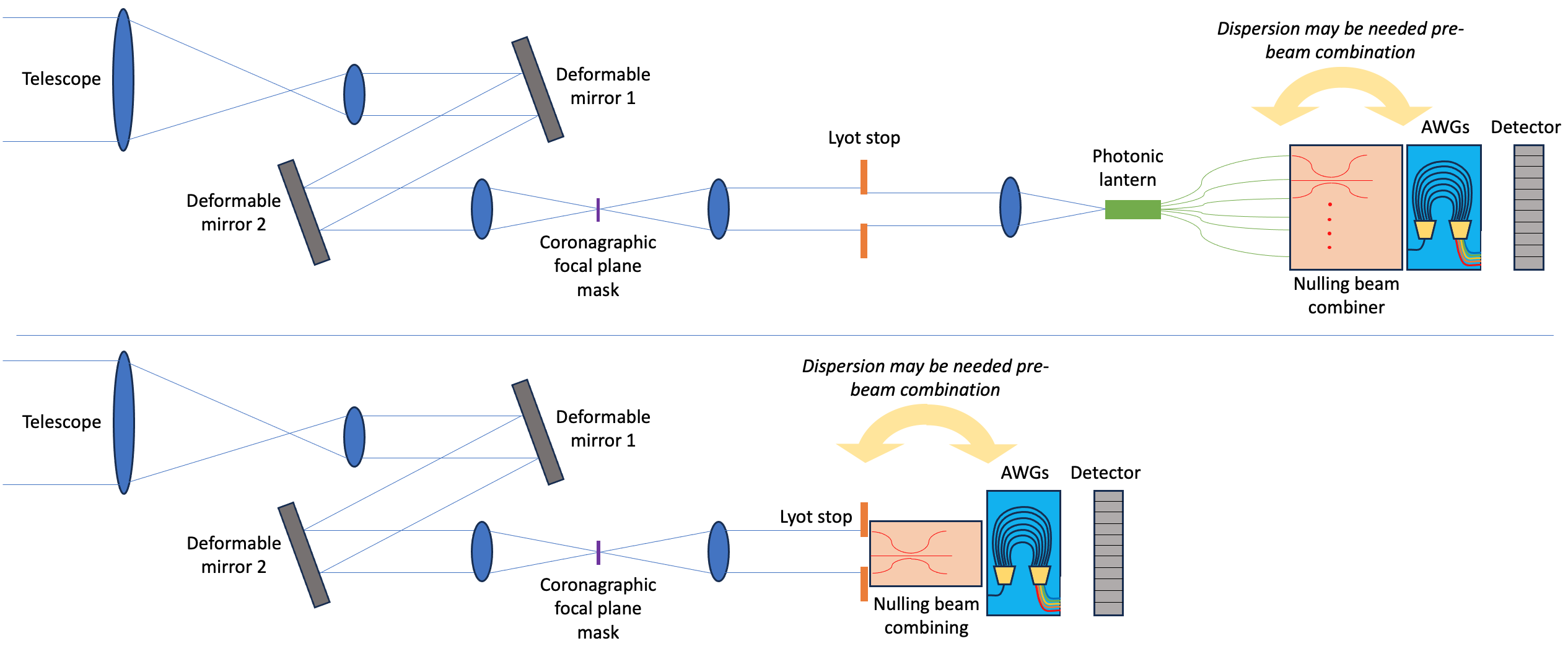}
\end{center}
\caption[example] 
   { \label{fig:architectures} 
Two potential photonic-based planet detection and characterization instrument architectures for the HWO. Light is captured by the telescope, and then bounces off the 2 deformable mirrors used for wavefront control before passing through a coronagraph (shown as a focal plane mask) before being filtered at the downstream Lyot stop. (Top) The light is injected into a photonic lantern and routed to a downstream beam combiner before being dispersed in AWGs and sent to a detector. (Bottom) Light is captured at the pupil plane immediately following the Lyot stop by a 3D pupil remapper and beam combiner chip. The outputs are dispersed and sent to a detector.}
\end{figure}

There are numerous possible architectures that a photonic-based instrument could take. However, given the extreme contrast needed to detect and characterize and Earth-like planet around a sun-like star, its unlikely photonics will do it alone. A more likely scenario is to use a coronagraph at moderately high contrast (10$^{-6}$-10$^{-7}$ at 2 $\lambda/D$) and then use photonic components down stream to build on this. Several possible architectures are shown in Fig.~\ref{fig:architectures}.  

One architecture could consist of a GLINT like nuller injecting light in the same pupil plane as the Lyot stop of the coronagraph (see bottom panel of Fig.~\ref{fig:architectures}). Another possibility is to use the mode-selective lantern nuller in a downstream focal plane with a back end beam combiner chip optimized for Kernel nulling (see top panel in Fig.~\ref{fig:architectures}). This approach has the added advantage that that mode-selective lantern is a passive component offering some starlight suppression without the need for sensing and active control. The Kernel nuller however will need some active control, which again will be challenging at the low photon rates expected after such extreme starlight suppression. Another approach would be to segment the pupil into two apertures, inject each into individual mode selective photonic lanterns and then combine the outputs in a chip to realize a double Bracewell architecture~\cite{Tuthill_NIH_2022}. 

In addition to the coronagraph and photonics, the use of masks (vortex, phase knife, etc) and/or phase plates (applied to the deformable mirrors or a separate plate) need to also be considered in the architecture.  

Photonic spectrographs will be critical to characterize the planet. However, if the beam combiner chips can not be made to provide deep nulls over broad bandwidths, then AWGs might need to be used immediately after light collection. In this way the spectral channels are split early on and then downstream beam combination is conducted on a spectral channel by spectral channel basis with narrower overall band passes. This is similar to the layout of the prototype spectrum shaper on a chip~\cite{Jovanovic_FLF_2022}. This increases the complexity of the circuit as well as the total number of degrees of freedom, which should ultimately allow for a broader null to be achieved. 

The challenge's that need to be addressed include:
\begin{itemize}
    \item Defining if and what sort of coronagraph is ideally suited to work in tandem with a photonic nuller. 
    \item Understanding the chromatic behavior of the various photonic nulling options and if pre-dispersion can be used as a viable pathway to broaden the null and 
    \item Investigating how to control and calibrate photonic components used with low photon rates far downsteam in the starlight suppression system. 
\end{itemize}

\section{Summary}
We have outlined how the functionalities of photonic technologies could be exploited to detect and characterize terrestrial planets in the habitable zones of sun-like stars. From the ground, giant segmented mirror telescopes could be used to detect similar planets around cooler M stars using similar technical solutions. Technologies like LFCs are already at technology readiness level (TRL) 9 and are being widely used at ground based observatories. Nulling and wavefront control technologies range from TRL 2-5 depending on the approach. There is a lot of work to be done to advance all these concepts to a sufficient readiness level to be able to properly evaluate their potential for application to the HWO. 

This work closely follows the 2023 astrophotonics roadmap~\cite{Jovanovic_ARP_2023}. For more details, please consult the roadmap.

\acknowledgments % equivalent to \section*{ACKNOWLEDGMENTS}       
 
This work has been supported by the National Science Foundation under Grant No. 2109231. Y.X acknowledges support from the National Science Foundation Graduate Research Fellowship under Grant No. 1122374. This work was supported by the Wilf Family Discovery Fund in Space and Planetary Science, funded by the Wilf Family Foundation. This research was carried out in part at the California Institute of Technology and the Jet Propulsion Laboratory under a contract with the National Aeronautics and Space Administration (NASA). Support for P Gatkine was provided by NASA through the David \& Ellen Lee Prize Postdoctoral Fellowship and NASA Hubble Fellowship Grant HST-HF2-51478.001-A awarded by the Space Telescope Science Institute, which is operated by the Association of Universities for Research in Astronomy, Incorporated, under NASA Contract NAS5-26555.

% References
\bibliography{report} % bibliography data in report.bib
\bibliographystyle{spiebib} % makes bibtex use spiebib.bst

\end{document}